# THE FLOWING SYSTEM GASDYNAMICS
## Part 4: Influence of the incident flow velocity on the outflow velocity out of flowing element

S.L Arsenjev, I.B. Lozovitski[1], Y.P.Sirik

*Physical-Technical Group*
*Dobroljubova street 2, 29, Pavlograd, Dnepropetrovsk region, 51400 Ukraine*

It is shown, that the introduction in Saint-Venant -Wantzel's formula radicand of the free component in the form of quadrate of velocity of the incident gas flow on inlet of flowing element that was offered in the end of the XIX century and is remained till now it is physically inadequate. It is shown that this way is doubtful for hydromechanics and it absolutely isn't reasonably for gasdynamics. Physically exact form of writing of static head law for flowing element and relevant to this, the modern form of Saint-Venant - Wantzel's formula are submitted ad-hoc. The obtained expressions allow physically correctly to take into account both combined and isolated influence of pressure drop applying to the flowing element and velocity of incident gas flow on flowing element, on quantity of the outflow velocity of gas stream out of this flowing element, system. The obtained expressions are valid for a subsonic velocity of the incident gas flow. The particular expressions are obtained for liquid.

## Nomenclature

$\gamma_{en}$    weight density of incident flow to inlet of flowing element

$\lambda$    coefficient of hydraulic friction

$g$    acceleration of gravity

$l$    current length of stream

$L$    general length of flowing element

$K = \left(1 - \dfrac{l}{L}\right)$ relative current length of stream in flowing element

$\bar{L}$    general caliber length of flowing element, $L/D$

$\zeta_{in}$    coefficient of local hydraulic resistance for inlet into flowing element

$\zeta_{ex}$    coefficient of local hydraulic resistance for outlet from flowing element

$p_0, p_h$    quantities of pressure before inlet and on outlet of flowing element accordingly

$p_{st}(l)$    current quantity of static head in stream

$V_{en}$    velocity of the incident flow on inlet of flowing element

## 1 Introduction

The problem solution on the combined action of pressure drop applying to the flowing element and velocity of incident flow on the inlet of flowing element is general problem for description of the gas and liquid flow motion.

This problem was solved in hydromechanics by introduction in the radicand of the Torricelli-Galilei-Borda-Du Buat (TGBD) formula of free component in the form of velocity quadrate of the incident flow on the inlet of the flowing element [1]. Thus, the first component of radicand of the TGBD formula allows for pressure drop applying to the flowing element, the second component allows for the incident flow velocity on flowing element and then the square root from this total with allowance for the velocity coefficient determins the flow rate of fluid on outlet of the flowing element. The circumscribed formula exists in such form in hydromechanics from the beginning of the XIX century till now. In the end of the XIX century, this way was transferred into gasdynamics: the free component in the form of velocity quadrate of incident gas flow on the inlet of the flowing element was also brought into the radicand of the Saint-Venant - Wantzel formula (SVW) [2,3]. The velocity coefficient is present in the front of root in this case as well as in hydromechanics. The empirical-speculative character of the circumscribed approach is obvious in hydromechanics and especially in the gasdynamics. For example, if only the velocity of incident gas flow ("*flying pipe*" version) acts on the inlet of the flowing element without pressure drop, the outlet velocity out of the flowing element without of account of the velocity coefficient will be equal to velocity of incident flow. In this case, the strange situation is arisen: the friction does not influence upon the moving gas stream in flowing element. This is not the gas stream, this is physically empty medium under title an ideal fluid. The problem is: how does the given situation, with a stale more than 175 years, to solve physically adequately and mathematically definitely?

---

[1] Phone: (38 05632) 38892, 40596
E-mail: loz@inbox.ru

## 2 Approach

The overcoming of the problem of contact interaction of fluid medium with streamline surface has allowed to find the unitized expression for allocation of static head of the gas or liquid stream along the length of flowing element in the form of law of static head [4, 5]. Side by side with it SVW formula was also led to the final physically correct form [6]. If these expressions have the property of adequate reflection of a physical reality, then they should allow to take into account all multiplicity of mechanical actions on fluid stream. Including, they should allow to take into account the combined action of pressure drop and velocity of incident flow on the flowing element and moreover without any tales like of the velocity and flow rate coefficients and polytrope [7] and other well-known "betterments."

## 3 Solution

Despite of the brevity of the approach, the solution of the problem appears rather simple. There is enough, the velocity head of incident flow on inlet of the flowing element to add to $p_0$ in the law of static head and the solution is reached.

So, velocity head of incident flow:

$$p_{en} = \frac{\gamma_{en} \cdot V_{en}^2}{2g} \qquad (1)$$

In this case, the law of static head for gas stream in the flowing element:

$$p_{st}(l) = ((p_0 + p_{en}) - p_h) \times \\ \times \frac{\lambda \overline{L} K_l}{1 + \lambda \overline{L} K_l + \zeta_{in} + \zeta_{ex}} + p_h \qquad (2)$$

At substitution of the $p_{st}(l)$ to the modern form of SVW formula [6] it will accept a form:

$$V_{ex} = \left(\frac{2}{k-1} kgRT_0 \left(1 - \left(\left(1 + \frac{p_{en} - p_h}{p_0}\right) \times \right.\right.\right. \\ \left.\left.\left. \times \frac{\lambda \overline{L}}{1 + \lambda \overline{L} + \xi_{in} + \xi_{ex}}\right)^{\frac{k-1}{k}}\right)\right)^{\frac{1}{2}} \qquad (3)$$

The law of static head for the liquid stream in flowing element, with taking into account that the expression for the velocity head of incident liquid flow has the same form as for gas flow, looks like:

$$p_{st}(l) = ((p_0 + p_{en}) - p_h) \times \\ \times \frac{\lambda \overline{L} K_l}{1 + \lambda \overline{L} + \zeta_{in} + \zeta_{ex}} + p_h \qquad (4)$$

At the same time, TGBD formula [4] will accept the form appropriate for hydromechanics:

$$V = V_{ex} = \left(2 \frac{g}{\gamma} \frac{p_0 + p_{en} + p_h}{1 + \lambda \overline{L} + \xi_{in} + \xi_{ex}}\right)^{\frac{1}{2}} \qquad (5)$$

during the outflow of liquid stream under the reposing liquid level. Also it will look like:

$$V = V_{ex} = \left(2 \frac{g}{\gamma} \frac{p_0 + p_{en} + p_h}{1 + \lambda \overline{L} + \xi_{in}}\right)^{\frac{1}{2}} \qquad (6)$$

during the outflow of liquid stream to an atmosphere.

We is obtaining "*flying pipe*" variant at activity of the velocity of the incident flow alone on inlet of the flowing element and in absence of pressure drop. For this case, the law of static head has the appearance of:

$$p_{st}(l) = ((p_{en} + p_h) - p_h) \times \\ \times \frac{\lambda \overline{L} K_l}{1 + \lambda \overline{L} K_l + \zeta_{in} + \zeta_{ex}} + p_h \qquad (7)$$

or

$$p_{st}(l) = p_{en} \times$$
$$\times \frac{\lambda \overline{L} K_l}{1 + \lambda \overline{L} K_l + \zeta_{in} + \zeta_{ex}} + p_h \quad (8)$$

At substitution of this expression in the modern form of writing of SVW formula, the latter will accept the view:

$$V_{ex} = \left( \frac{2}{k-1} kgRT_0 \left( 1 - \left( \frac{p_{en}}{p_h} \times \right.\right.\right.$$
$$\left.\left.\left. \times \frac{\lambda \overline{L}}{1 + \lambda \overline{L} + \xi_{in} + \xi_{ex}} \right)^{\frac{k-1}{k}} \right) \right)^{\frac{1}{2}} \quad (9)$$

in case of subsonic velocity of incident gas flow.

## 4 Final remarks

Thus, the problem about the physically adequate and mathematically precise taking into account of influence of velocity of incident gas flow on outlet velocity is resolved for two practically important cases: a combined action of pressure drop and velocity of incident gas flow and isolated action of velocity of incident gas flow. The particular solutions are given for fluid flow

[1] Handbook for industrial engineer, Vol.2, The State Publishing House Mashgiz, p.478, 1955

[2] W. Schule, "Technische Warmemechanik," Verlag von Julius Schpringer, Berlin, 1909

[3] E.Gertz, G.Kreinin, "Calculation of the pneumatic drives," Handbook, The State Publishing House Mashinostroenie, Moscow, 272p, 1975

[4] S.L. Arsenjev, I.B. Lozovitski, Y.P. Sirik, "The flowing system gasdynamics. Part 1: On static head in the pipe flowing element,"
**http://arXiv.org/abs/physics/0301070** , 2003

[5] S.L. Arsenjev, I.B. Lozovitski, Y.P. Sirik, "The flowing system gasdynamics. Part 2: Euler's momentum conservation equation solution,"
**http://arXiv.org/abs/physics/0302020,** 2003

[6] S.L. Arsenjev, I.B. Lozovitski, Y.P. Sirik, "The flowing system gasdynamics. Part 3: Saint-Venant –Wantzel's formula modern form"
**http://arXiv.org/abs/physics/0302038**, 2003

[7] G.A. Zeuner, "Technische Thermodynamik," Vol.2, Leipzig, 1900